\documentclass[twocolumn,twocolappendix]{aastex631}

\def\eg{{\it e.g.}}
\def\etal{{\it et al.}}

\def\ie{{\it i.e.}}

\def\NB{{\it NB.}}

\def\pmb#1{\setbox0=\hbox{$#1$}%
  \kern-0.25em\copy0\kern-\wd0
  \kern.05em\copy0\kern-\wd0
  \kern-0.025em\raise.0433em\box0}
\def\spmb#1{\setbox1=\hbox{${\scriptstyle #1}$}%
  \kern-0.25em\copy1\kern-\wd1
  \kern.05em\copy1\kern-\wd1
  \kern-0.025em\raise.0433em\box1}

\long\def\Ignore#1{\relax}

\def\spose#1{\hbox to 0pt{#1\hss}} 
\def\gtlt{\mathrel{\spose{\lower.5ex\hbox{$\mathchar"13E$}}
     \raise.5ex\hbox{$\mathchar"13C$}}}
\usepackage{color}
\definecolor{red}{rgb}{0.7,0.1,0.1}
\definecolor{blue}{rgb}{0.2,0.2,0.8}
\definecolor{green}{rgb}{0.1,0.6,0.1}

\submitjournal{ApJ}

\graphicspath{{./}{figures/}}

\begin{document}

\title{The quest for a stable disk}
\shorttitle{Disk stability}

\author{J. A. Sellwood}
\affiliation{Steward Observatory, University of Arizona \\
933 Cherry Avenue \\
Tucson, AZ 85722, USA}

\author{R. G. Carlberg}
\affiliation{Department of Astronomy and Astrophysics \\
University of Toronto \\
Toronto, ON M5S 3H4, Canada}

\shortauthors{Sellwood \& Carlberg}

\begin{abstract}
  The majority of disk galaxies manifest spirals and/or bars that are
  believed to result from dynamical instabilities.  However, some
  galaxies have featureless disks, which are therefore inferred to be
  dynamically stable.  Yet despite many years of effort, theorists
  have been unable to construct realistic models of galaxy disks that
  possess no instabilities and therefore could remain featureless.
  This conclusion has been reached through simulations for the most
  part, some of which have been confirmed by linear stability
  analyses. Toomre claimed that the Mestel disk, embedded in an equal
  mass halo, to be a notable counter-example, but his prediction of
  stability could not be reproduced in simulations due to complicated
  non-linear effects that caused secular growth of Poisson
  noise-driven disturbances until strong features emerged.  Here we
  revisit this issue and show that simply eliminating the most nearly
  circular orbits from Toomre's disk model can inhibit troublesome
  secular growth.  We also present both 2D and 3D simulations of
  particle disks that remain featureless for over 50 orbit periods. We
  report that spiral evolution naturally depletes circular orbits and
  that the radial velocity distribution in the featureless disks of S0
  galaxies should have negative kurtosis.
\end{abstract}

\keywords{Spiral galaxies (1560) --- Galaxy structure (622) --- Galaxy dynamics (591) --- Galaxy evolution (594)}

\section{Introduction} \label{sec.intro}
Because the stars of a disk galaxy generally move about the center in
an ordered streaming flow with a small degree of random motion, we
find that many disk galaxy models are beset by gravitationally-driven,
collective instabilities.  Bar-forming modes \citep{Hohl71, Kaln78}
are the most persistent, while others are edge modes \citep{Toom81,
  PL89}, groove modes \citep{SL89, SK91}, and buckling modes
\citep{Toom66, MS94}.  We observe that most disk galaxies manifest
spiral arms and more than half have either weak or strong bars
\citep{Erwi18}, which are believed to result from instabilities
\citep[see][for a review]{SM22}. However, there are some that have
neither, which challenge dynamicists to construct somewhat realistic
models of disk galaxies that are completely stable.  Not only is this
a challenge in its own right, but such models would be valuable to
study responses to perturbations without the complication of intrinsic
instabilities.

Yet, to our knowledge, only two types of stable, rotationally
supported stellar disk models are known.  The first stems from the
classic study of uniformly rotating 2D disks by \citet{Kaln72}, who
was able to derive the full spectrum of normal modes of a class of
models that rotated at a uniform rate $\Omega$ that was slower than
the circular rate.  All his single $\Omega$ models possessed some
instabilities, but he was also able to construct composite models by
superposing variously weighted single $\Omega$ models and found two
cases that possessed no in-plane instabilities whatsoever.
Unfortunately, not only is uniform rotation unrealistic, but the same
models are unstable to buckling modes \citep{Poly77} when motion out
of the plane is permitted.

The other case, presented by \citet{Toom81}, is the flattened
half-mass Mestel disk which we fully describe in \S\ref{sec.Mestel}
below.  Toomre was unable to find any instabilities in his linearized
stability analysis, and therefore claimed that the disk should be
stable.  Note that the Mestel disk is also a 2D model that would
suffer from buckling instabilities \citep{MS94} if motion outside the
disk plane were allowed, but buckling modes are suppressed in disks of
finite thickness if the vertical velocity dispersion of the stars is
$\ga 1/3$ that of the radial dispersion \citep{Toom66, Arak87}.

\citet[][hereafter S12]{Sell12} tested Toomre's prediction, and
confirmed that his larger-$N$ simulations of the initial model did not
possess any {\em linear} instabilities.  However, he found
unexpectedly that simulations with even very large numbers of
particles, up to 500 million, always manifested secular growth in two
stages.  At first, the leading components of the inevitable shot noise
were amplified as they sheared to trailing.  The amplified disturbance
extracts energy from the gravitational well of the disk through
outward angular momentum transport, which must later be absorbed at
the Lindblad resonances \citep{LBK, Mark74}.  The energy absorbed by
the resonant particles is second order in the disturbance amplitude,
changing, or ``scratching,'' the DF very slightly.  The scratch is
localized in angular momentum and increases the random energy of the
resonant stars thereby creating a deficit of low-$E_{\rm rand}$ stars.
Subsequent amplified noise is affected by previous scratches, causing
responsiveness of the disk to grow gradually, and the general level of
activity also to rise.  Eventually, the scratches become deep enough
to seed a true groove-type instability, at which point amplitude
growth becomes exponential, leading to the appearance of a finite
amplitude bar.  \citet{FP15} were able to reproduce Sellwood's
findings using quasi-linear theory, even to predict, from Sellwood's
input noise spectrum, the frequency of the instability that eventually
emerged \citep{RFP19}.  There seems no reason that this sequence of
events can be avoided no matter how many particles are employed, and
therefore Toomre's prediction of linear stability is undermined by
non-linear resonant scattering.

\citet{SC23} reviewed the extensive literature on bar instabilities,
which we do not repeat here.  The conclusions from their survey were
that heavy halos do suppress bar instabilities in cool disks, but only
to unleash multi-arm spiral instabilities in their place.  Bar
instabilies can also be quelled by raising $Q$, however no analytical
requirement for non-axisymmetric stability is known.  A survey of then
available numerical results led \citet{AS86} to propose that ``$Q
\simeq 2-2.5$ at all radii as a sufficient criterion for
non-axisymmetric stability''.  However, as noted above, high radial
velocity dispersion in disks mandates substantial vertical thickness
to inhibit buckling instabilities, requiring an unrealistic degree of
puffiness if full-mass disks are to be stable.

Here we continue the search for a more nearly stable disk model, by
showing that the troublesome secular growth in the Mestel disk
reported by S12 can be suppressed by a minor modification to the DF.
Since the secular growth is caused by groove-type behavior, we must
first highlight one property of groove modes that has not previously
been emphasized, but which will be important to the solution we
propose.

\section{Mestel disk model} \label{sec.Mestel}
The razor-thin Mestel disc used in the studies by \citet{Zang76} and
\citet{Toom81} is characterized by a constant circular speed $V_0$ at
all radii, and has the mid-plane potential $\Phi(R) = V_0^2
\ln(R/R_0)$ where $R_0$ is a reference radius that defines the zero of
the potential.  The axisymmetric surface density, $\Sigma_0(R) = V_0^2
/ (2\pi G R)$, would self-consistently yield the appropriate central
attraction for centrifugal balance for $0 \leq R \leq \infty$.
\citet{Toom77} and \citet{BT08} give the DF for this disk that has a
Gaussian distribution of radial velocities with spread $\sigma_R$ at
all radii:
\begin{equation}
f(E,L_z) = \cases{ F \, (L_z/R_0V_0)^q e^{-E/\sigma_R^2} & $L_z>0$ \cr
           0 & otherwise, \cr}
\label{eq.DF}
\end{equation}
where $q = V_0^2/\sigma_R^2 - 1$, and the normalization constant is
\begin{equation}
F = {1 \over G R_0} { (q/2 + 0.5)^{q/2+1} \over \pi^{3/2}(q/2-0.5)!}.
\end{equation}

\citet{Zang76} applied a central cutout to the disk, described below,
in order to avoid difficulties with particles having near infinite
frequencies as $L_z \rightarrow 0$, which he replaced by bulge-like
rigid matter to preserve the central attraction $a_R=-V_0^2/R$ at all
radii.  His linear stability analysis of this centrally cut-out, but
otherwise full-mass, disk revealed a persistent lop-sided ($m=1$)
instability no matter how gentle the central cutout.  Therefore
\citet{Toom81} studied the stability of the same cutout model but
with the active mass halved to suppress the lop-sided mode, with the
removed mass replaced by rigid halo-like matter to preserve the disk
equilibrium.

While maintaining the centripetal acceleration $a_R = -V_0^2/R$
everywhere, \citet{Toom81} multiplied the DF (eq.~\ref{eq.DF}) by
the double taper function
\begin{equation}
T(L_z) = 
\left[ 1 + \left( {R_0V_0 \over L_z} \right)^\nu \right]^{-1}
\left[ 1 + \left( {L_z \over R_1V_0} \right)^\mu \right]^{-1},
\label{eq.tapers}
\end{equation}
to create the central cutout having mean radius $R_0$, as well as an
outer taper having mean radius $R_1$ to limit the radial extent of the
otherwise semi-infinite disk. He chose $q=11.44$ or $\sigma_R =
0.284V_0$ to ensure $Q \rightarrow 1.5$ outside the inner cutout of
the half-mass Mestel disk.  Setting the taper indices $\nu=4$ and
$\mu=5$ yielded an idealized, smooth disc model that \citet{Toom81}
claimed possessed no small amplitude unstable modes.  Here we choose
$R_1 = 11.5R_0$, as did S12, and further limit the radial extent of
the disc by an energy cut-off to eliminate particles having sufficient
energy to pass $R=20R_0$.  We shall refer to this DF, which was
employed in S12, as our ``standard DF'' and we adopt units such that
$G=V_0=R_0=1$.

\begin{figure}
\begin{center}
\includegraphics[width=.99\hsize,angle=0]{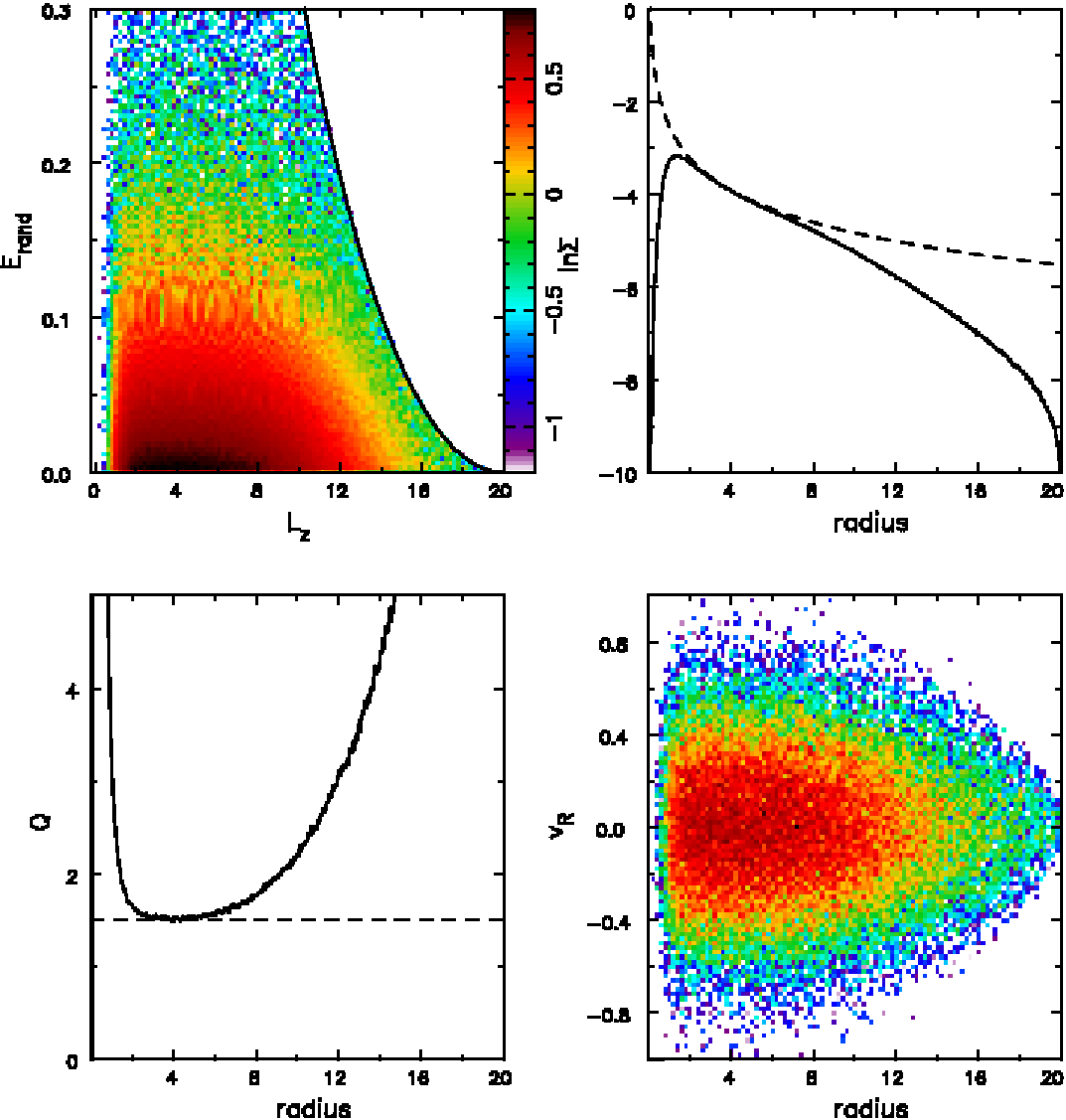}
\end{center}
\caption{Properties of our standard DF for in the doubly-tapered,
  half-mass Mestel disk.  Top left: The particle distribution in
  $(L_z,E_{\rm rand})$; the solid line is the upper bound on energy
  imposed by restricting particle orbits to not stray beyond $R=20$.
  Top right: The solid curve gives the active surface density of the
  disk, the dashed curve is what it would be without the tapers.
  Bottom left: The solid curve shows the radial variation of $Q$ while
  the dashed curve shows the value $Q$ would have in a disk lacking
  tapers.  Bottom right: the distribution of radial velocities, for
  which $\sigma_R\simeq 0.284$ in the self-similar disk, but is lower
  in the very inner and outer parts due to the tapers.}
\label{fig.5654}
\end{figure}

Figure~\ref{fig.5654} presents some properties of our standard DF with
the tapers and energy restriction applied to the function
(\ref{eq.DF}). The top left panel presents the density of particles in
$(L_z,E_{\rm rand})$ space, where
\begin{equation}
  E_{\rm rand} = E - E_c(L_z),
\end{equation}
is the excess of energy over that of a circular orbit of $L_z$ and the
right-hand boundary, delineated by the solid curve, marks the
restriction that no particle has sufficient energy to pass $R=20R_0$.
Note that the typical random energy of a particle having $v_R
=\sigma_R$ is $E_{\sigma_R} = \sigma_R^2/2 \simeq 0.04$.  The
surface density profile of the active disk is marked by the solid line
in the top right panel, which closely matches the surface density of
the untapered, half-mass disk (dashed curve) over the radial range $2
\la R \la 8$.  The untapered disk would be self-similar with $Q=1.5$
and $\sigma_R\simeq 0.284$ at all radii, but the radial velocities
(bottom right panel) of the active particles are limited by both the
inner and outer tapers.  The solid curve in the bottom left panel
shows the radial variation of $Q$ measured from the surface density
and velocity distribution of the active particles.

\begin{table}
\caption{Default numerical parameters for 2D} 
\label{tab.DBHpars}
\begin{tabular}{@{}ll}
Grid points in $(r, \phi)$ & 220 $\times$ 256 \\
Grid scaling & $R_0 = 10$ grid units \\
Active sectoral harmonic & $m=2$ \\
Plummer softening length & $\epsilon = R_0/10$ \\
Number of particles & $5 \times 10^6$ \\
Largest time-step & $0.08R_0/V_0$ \\
Radial time step zones & 5 \\
Guard radii & 4 \\
\end{tabular}
\end{table}

\section{Simulation code} \label{sec.code}
We select particles from the adopted DF using the procedure described
in \citet{Sell24}, place them in a plane at random azimuths and
compute the mutual attractions of the particles using a 2D polar mesh.
This code, which has been widely used and tested, is described in
detail in \citet{Sell14}; briefly the particles move subject to forces
from other particles that are interpolated from the grid.  We adopt
the parameters listed in Table~\ref{tab.DBHpars}.

In the simulations we report here, the central attraction is always
$a_R=-V_0^2/R$, which therefore includes the attraction of all rigid
material as well as that of the mobile particles and, unless otherwise
stated, only the $m=2$ sectoral harmonic contributes to the
non-axisymmetric terms.  We employ block time steps that are decreased
by factors of 2 in each radial zone and are further subdivided in a
succession of guard zones near the bulge-dominated center where
self-gravity of the disk is not updated at every sub-step.

As usual, we measure non-axisymmetric distortions of the distribution
of the $N$ particles using an expansion in logarithmic spirals:
\begin{equation}
A(m,\gamma,t) = {1 \over N}\sum_{j=1}^N \, \exp[im(\phi_j + \tan\gamma \ln R_j)],
\label{eq.logspi}
\end{equation}
where $(R_j,\phi_j)$ are the polar coordinates of the $j$th particle
at time $t$, $m$ is the sectoral harmonic, and $\gamma$ is the
(radially constant) angle of the spiral component to the radius
vector, which is the complement to the spiral pitch angle.

\section{Groove modes} \label{sec.groove}
Following \citet{SB02}, we multiply our standard DF
(Fig.~\ref{fig.5654}) by a Lorentzian function of $L_z$ to create an
axisymmetric groove
\begin{equation}
{\cal G}(E,L_z) = H(E_{\rm rand}) \left[ 1 + {\beta w_{L_z}^2 \over
    (L_z-L_*)^2 + w_{L_z}^2} \right],
\label{eq.groove}
\end{equation}
and generally choose the parameters: mean $L_*=6.5R_0V_0$, depth
$\beta=-0.4$, and width $w_{L_z}=0.1R_0V_0$.

In the study by \citet{SB02}, $H$ was set to unity and the groove
extended over all bound values of $E_{\rm rand}$.  But more recent
experience \citep[\eg][]{SC19} suggests that grooves among the low
$E_{\rm rand}$ particles are the principal source of instability.
Therefore we here study how groove instabilities are affected when we
restrict the extent of the groove to low, or high, values of $E_{\rm
  rand}$ by setting $H = 0$ or 1, depending upon whether
\begin{equation}
  E_{\rm rand} \gtlt E_{\rm rand, crit},
\end{equation}
for various values of $E_{\rm rand, crit}$.

\begin{figure}
\begin{center}
\includegraphics[width=.99\hsize,angle=0]{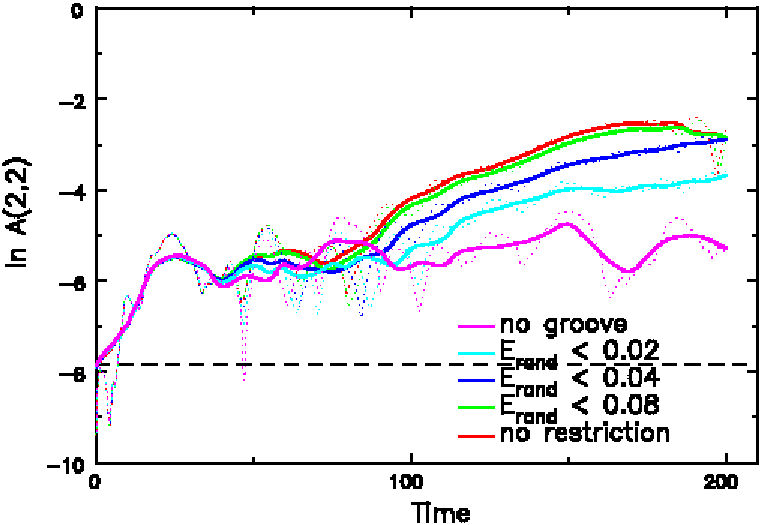}
\end{center}
\caption{The dotted lines report the time evolution of the $m=2$,
  $\tan\gamma=2$ logarithmic spiral in a series of simulations having
  grooves present over the indicated range of $E_{\rm rand}$ in each
  case while the heavily smothe values are marked by the solid curves.
  As the height of the groove in $E_{\rm rand}$ is increased from zero
  (magenta curve), the vigor of the instability increases until there
  is no upper $E_{\rm rand}$ bound on the groove (red curve).  Notice
  that the instability is only mildly less vigorous when the groove is
  restricted to $E_{\rm rand} < 2E_{\sigma_R}$ (green curve) than when
  the groove is present for all $E_{\rm rand}$ (red curve).}
\label{fig.cool}
\end{figure}

Fig.~\ref{fig.cool} presents results from several simulations, two of
which have been published before.  The magenta line is from a
simulation with no groove and uses identically the same initial DF,
except for the random draw, as the $N=5$M particle run in S12.  The
raw measure from eq.~(\ref{eq.logspi}) is plotted as the dotted line
in each case and the heavily smoothed data are indicated by the solid
lines.  Fig.~2 of S12 used a different procedure to smooth the $m=2$
amplitude, but the behavior was qualitatively the same as shown here.
The measured amplitude is initially that expected from randomly placed
particles, which is marked by the horizontal dashed line, but rises
rapidly by about a factor of ten as the particle distribution becomes
polarized \citep{TK91} due to the mutual attractions of the disk
particles.  Subsequently, the amplitude of the dotted magenta line
fluctuates about the mean (solid line), as randomly excited transient
spirals in different parts of the disk overtake one another causing
alternating constructive and destructive interference in the measured
amplitude.  The duration of the simulations shown in
Fig.~\ref{fig.cool} is too short to reveal the secular rise reported
by S12, but it becomes apparent later -- see \S\ref{sec.nosecular}.

The red line in Fig.~\ref{fig.cool} is from the model having a strong
groove instability, that was employed by \citet{SB02}.  The groove had
the factor $H = 1$ in eq.~(\ref{eq.groove}) and therefore extended
over all random energies.  Although a narrow feature in $L_z$, random
motion blurs the groove in radius, and since it cuts away just 1.1\%
of the disk mass, its effect on the $\Sigma$- and $Q$-profiles is
barely detectable.  The only differences are that here we employ half
the groove width and 5M particles, whereas \citet{SB02} used 6M.  The
groove instability causes the red line to emerge from the noise around
$t \simeq 70$ and to rise roughly exponentially to saturation near $t
\simeq 180$, a little earlier than in \citet{SB02} because of the
minor differences in the model.

\begin{figure}
\begin{center}
\includegraphics[width=.99\hsize,angle=0]{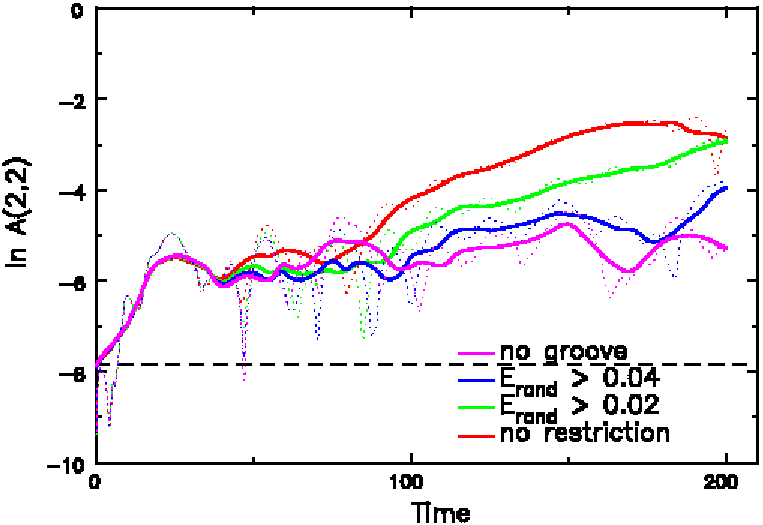}
\end{center}
\caption{As for Fig.~\ref{fig.cool}, but when the groove is present
  for higher random energies and not for lower.  The red and magenta
  curves are the same in both figures.}
\label{fig.hot}
\end{figure}

\subsection{Grooves among low $E_{\rm rand}$ particles only} \label{sec.cool}
Here we experiment by setting $H=1$ when $E_{\rm rand}<E_{\rm
  rand,crit}$ and $H=0$ otherwise, which creates a groove among the
particles of low $E_{\rm rand}$ only.

The results from a series of runs in which the groove is restricted in
this way to several values of $E_{\rm rand,crit}$ are also presented in
Fig.~\ref{fig.cool}.  The green curve manifests almost the full
instability when $E_{\rm rand,crit} = 0.08$, implying that yet higher
random energy particles take little part in the instability.  The
vigor of the instability decreases as the groove is restricted to
lower $E_{\rm rand}$. However, the cyan curve reveals that a mild
instability still persists when the groove is restricted to as little
as $E_{\rm rand,crit} = 0.02$.

\subsection{Grooves among high $E_{\rm rand}$ particles only} \label{sec.hot}
Fig.~\ref{fig.hot} shows the effect on the instability caused by a
groove in only the most eccentric orbits, while eliminating it among
the lowest $E_{\rm rand}$ particles.  Notice that the negligible
growth reported by the blue curve reveals that a groove among all the
particles having $E_{\rm rand} \ga 0.04$ has little destabilizing
effect.

The perhaps unsurprising conclusion from the new results presented in
Figs.~\ref{fig.cool} and \ref{fig.hot} is that groove instabilities
are provoked most strongly by deficiencies among the low-$E_{\rm rand}$
particles while groove-like deficiencies among those particles having
the most eccentric orbits have little effect on disk stability.

\begin{figure}
\begin{center}
\includegraphics[width=.99\hsize,angle=0]{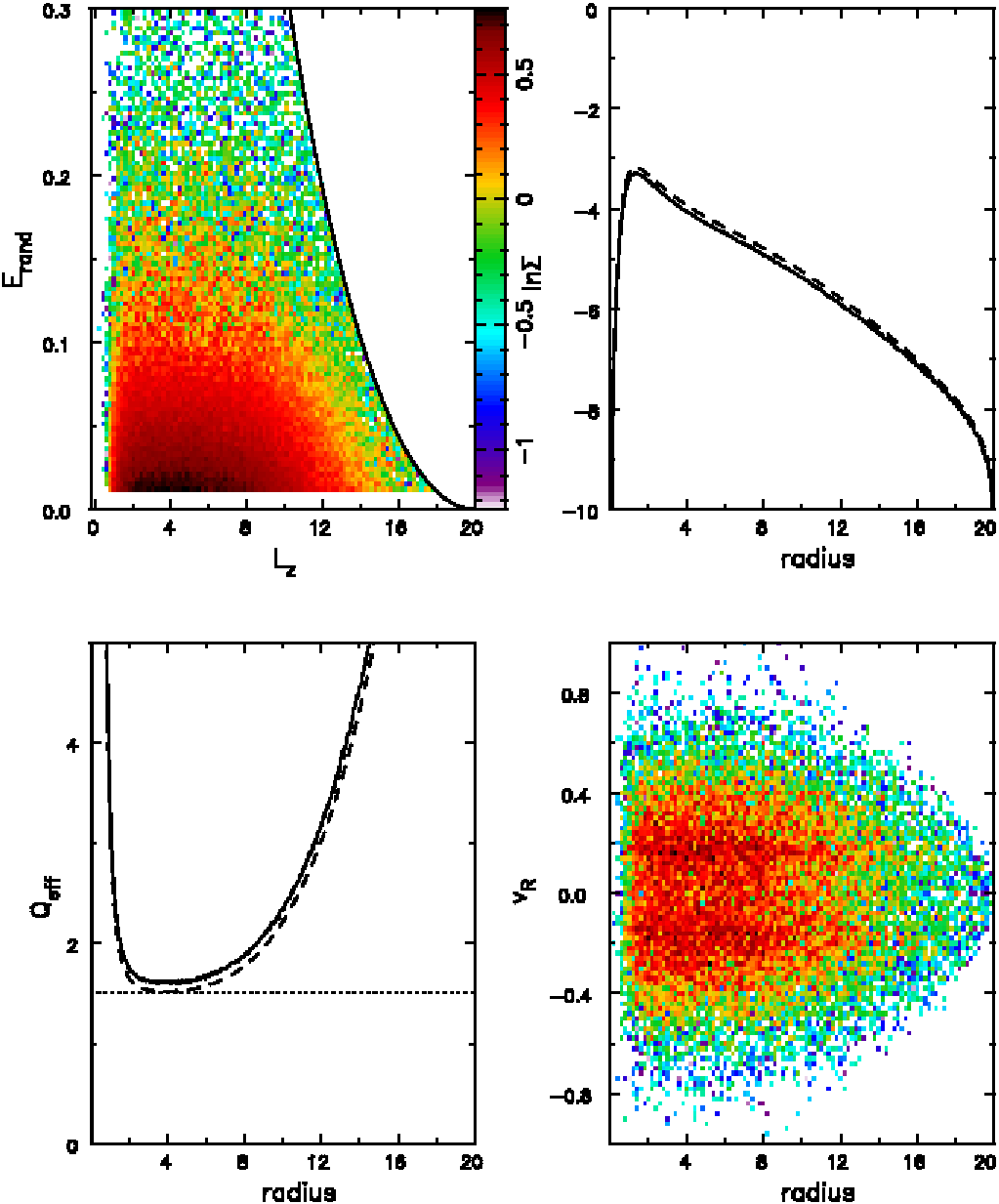}
\end{center}
\caption{Properties of the particle distribution when those having
  $E_{\rm rand}<0.01$ are excluded from the selection, as illustrated
  in the top left panel.  Eliminating the most nearly circular orbits
  slightly reduces the disk surface density (top right), raises
  $Q_{\rm eff}$, and makes the radial velocity distribution bimodal
  (bottom right).}
\label{fig.DF5682}
\end{figure}

\section{Inhibiting secular growth} \label{sec.nosecular}
With these results in hand, we now address the main challenge of
this paper.

S12 reported that simulations with particles drawn from our standard
DF, without initial grooves, did not remain featureless indefinitely.
While simulations with large $N$ did not, at first, support any
exponentially growing modes, confirming linear stability predicted by
\citet{Toom81}, all later manifested exponentially growing
disturbances.  It seemed that shot noise in the disk, amplified by the
shear, led to weak collective waves that caused very mild initial
resonance scattering, which ``scratched'' the DF.  The weak scratches
caused stronger collective responses in an ever increasing manner
until the changes to the DF were sufficient to trigger a genuine
instability that then grew exponentially to saturation.  Thus the
predicted linear stability of the model was undermined by non-linear
resonance scattering creating grooves in the DF that eventually led to
true instability.

However, we have shown above (\S\ref{sec.groove}) that groove
instabilities are driven most strongly by deficiencies among the
distribution of the low-$E_{\rm rand}$ particles, so if we were to set
up a disk in which the lowest $E_{\rm rand}$ particles were removed,
the recurrent cycle of scattering that led to the secular growth
reported by S12 might be suppressed.  In order to test this
hypothesis, we selected particles from the standard DF without any
grooves while excluding all that had low $E_{\rm rand}$.

\begin{figure}
\begin{center}
\includegraphics[width=.99\hsize,angle=0]{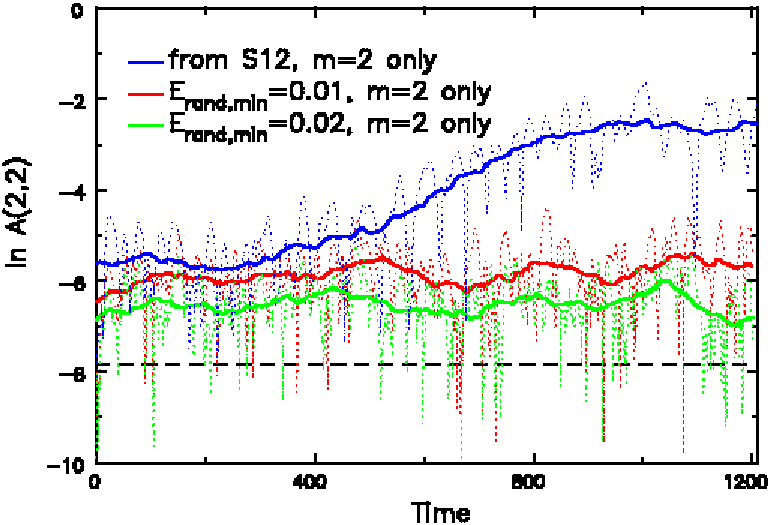}
\end{center}
\caption{The amplitude evolution of the $m=2$, $\tan\gamma=2$
  logarithmic spiral component in three simulations.  The blue curves
  are taken from the $N=5$M presented in S12 that employed the DF in
  eq.~(\ref{eq.DF}) without a groove. \NB\ Fig.~2 of that paper plots
  a different measure of the $m=2$ amplitude. The red line is from the
  simulation in which particles with $E_{\rm rand}<0.01$ were excluded
  from selection, as shown in Fig.~\ref{fig.DF5682}, while the green
  line is from a simulation that lacked particles with $E_{\rm
    rand}<0.02$ in which activity is even more subdued. Notice that
  secular growth is eliminated in both these cases}
\label{fig.Ermin}
\end{figure}

\begin{figure}
\begin{center}
\includegraphics[width=.99\hsize,angle=0]{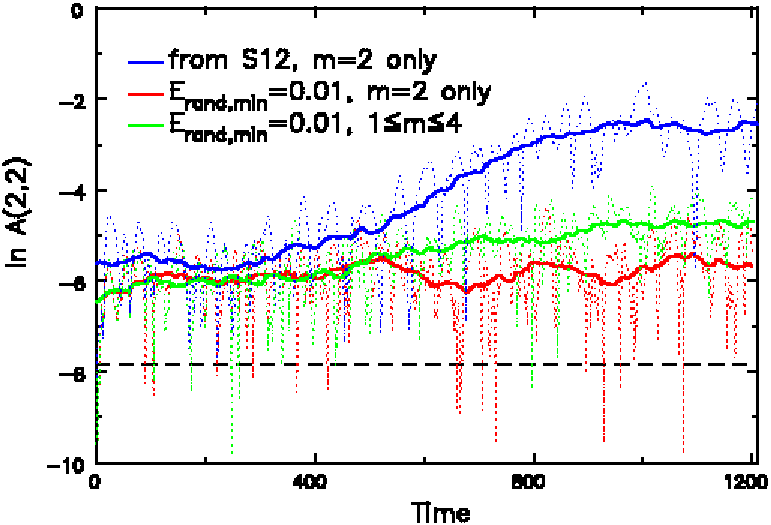}
\end{center}
\caption{The red and blue lines are the same as those
  fig.~\ref{fig.Ermin}, but the green line is a rerun of the same DF
  used for the red line with forces from $1\leq m \leq 4$ active.}
\label{fig.stable}
\end{figure}

\subsection{Excluding low $E_{\rm rand}$ particles}\label{sec.hotMes}
Starting from the standard DF that S12 employed, we eliminated all
particles having $E_{\rm rand}<0.01$ to obtain the DF shown in
Fig.~\ref{fig.DF5682}.  Notice that the surface density is reduced by
$\sim 11.6$\%, $Q$ is slightly increased, and the distribution of
radial velocities becomes bimodal, as shown in the top right, bottom
left, and the bottom right panels respectively.

Using this DF, we obtained the amplitude evolution of the largest
trailing bisymmetric component, the usual $m=2$, $\tan\gamma=2$, which
we compare in Fig.~\ref{fig.Ermin} with the result shown by the blue
line previously reported by S12 for the same $N$ that included
particles for all $E_{\rm rand}\ge 0$. (Note that a different measure
of the amplitude was plotted in Fig.~2 of the earlier paper.) These
simulations were run for six times longer than those presented in
Figs.~\ref{fig.cool} and \ref{fig.hot}. The red and green lines, which
manifest no secular growth, report the results when $E_{\rm rand} \ge
0.01$ and 0.02 respectively. The disk is a bit more lively with the
smaller lower bound on $E_{\rm rand}$, but even this case secular growth
was suppressed.

These results were obtained from simulations in which disturbance
forces were restricted to $m=2$, and it is of interest to
determine whether secular growth is also suppressed when disturbance
forces from multiple sectoral harmonics are active.  The green line in
Fig.~\ref{fig.stable} plots the $m=2$ amplitude from a simulation that
has the same DF given by the red line, but in which disturbance forces
from $1\leq m \leq 4$ were active.  The green line in this Figure is
slightly higher than the red, and the $m=1$ and $m=3$ amplitudes also
rose at a similar very slow rate.

It should be admitted, however, that the DF for this model with
negligible secular growth seems unnatural because the radial velocity
distribution is bimodal (see Fig.~\ref{fig.DF5682}).

\begin{table}
\caption{Default numerical parameters for 3D} 
\label{tab.3Dpars}
\begin{tabular}{@{}ll}
Grid points in $(r, \phi, z)$ & 220 $\times$ 256  $\times$ 125 \\
Vertical spacing of planes & $ 0.04R_0$ \\
Grid scaling & $R_0 = 10$ grid units \\
Active sectoral harmonics & $m=0 \& 2$ or  $m=0, 2, 3 \& 4$ \\
Softening length & $\epsilon = R_0/10$ \\
Number of particles & $5 \times 10^6$ or $5 \times 10^7$ \\
Largest time-step & $0.08R_0/V_0$ \\
Radial time step zones & 5 \\
Guard radii & 4 \\
\end{tabular}
\end{table}

\begin{figure*}
\begin{center}
\includegraphics[width=.3\hsize,angle=270]{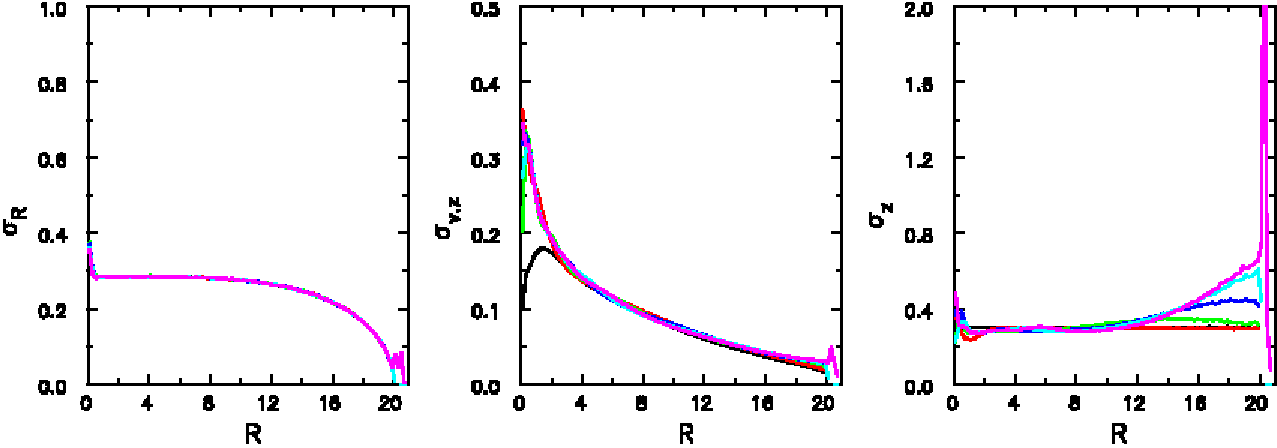}
\end{center}
\caption{The initial flaring of the 3D disk.  The left hand and middle
  panels show the radial and vertical velocity dispersions of the disk
  particles at $t=0$ (black) and at intervals of 6 time units in order
  red, green, blue, cyan, and magenta. The right hand panel shows the
  rms disk thickness increases at radii $R>10$ at the same times, but
  there is no further thickening after $t=30$.}
\label{fig.flare}
\end{figure*}

\section{Allowing 3D motion} \label{sec.3D}
All the results presented so far, as well as the prediction from
\citet{Toom81}, relate to a razor-thin disk.  However, real galaxies
have disks of finite thickness.  \citet{Toom82} describes a family of
exact axisymmetric, thickened disk models having similar properties to
those of the razor-thin Mestel disk, but they all pinch down to zero
thickness at the center, while being particularly puffy at
intermediate radii, and therefore do not make realistic 3D
generalizations of the Mestel disk.  Instead, we need to construct a
3D disk that resembles the half-mass Mestel disk model and which is
close to equilibrium.

\subsection{Setting up a thickened Mestel disk}
Our standard DF used by \citet{Toom81} (eqs.~\ref{eq.DF} \&
\ref{eq.tapers}) yields a suitable set of particles in a razor thin
disk, which we can spread vertically as a Gaussian in $z$ to create
the density distribution
\begin{equation}
\rho(R,z) = {\Sigma(R) \over (2\pi)^{1/2}z_0} \exp\left(- {z^2\over 2z_0^2}\right),
\end{equation}
with $z_0=0.3R_0$ being the disk scale height that is independent of
$R$.  But achieving both radial and vertical balance presents a number
of challenges.

The first challenge is that we must allow for forces that are not
confined to the disk plane.  With a 2D model, we could simply
establish equilibrium by setting the total central attraction to
$a_R=-V_0^2/R$, without any axisymmetric contribution from
self-consistent forces.  Notionally, we can think of half of this
force as arising from the attraction of the halo, because the disk
mass was halved, and the other half from a smooth razor-thin,
semi-infinite disk.  In 2D disks, it does not matter how the
attractions of these two components might vary out of the $z=0$ plane,
but it does for a thickened disk.

The thin disk particles have in-plane velocity components that create
an equilibrium model if the in-plane attraction were $-V_0^2/R$ at all
distances from the mid-plane.  ($R$ is the cylindical radius.) The
initial, vertically-spread, particle distribution is axisymmetric and
therefore their gravitational attraction computed through the grid
would be also.  We first use a 3D grid, with parameters in Table
\ref{tab.3Dpars}, to obtain the gravitational field of the disk
particles in their initial positions and compute their azimuthally
averaged central attraction $a_{R,\rm grid}(R,z)$ at every grid point.
We then preserve, in a separate copy of the grid, the value $-V_0^2/R
- a_{R,\rm grid}(R,z)$ and apply this correction to the
grid-determined forces from the particles at each grid point at each
time step to preserve the radial balance.

The grid-determined vertical attraction from the disk particles will
be considerably less than it should be because of softened gravity and
the fact that the tapers have removed a lot of mass.  However, we do
not have to try to reproduce the exact vertical forces; they need to
be strong enough only to require vertical velocties large enough to
inhibit buckling instabilities over most of disk.  We therefore add to
the vertical force the vertical component of the attraction from the,
assumed spherical, half-mass halo: $-V_0^2z/(2r^2)$, where $r$ is the
spherical radius and $z$ the distance from the mid-plane.

With the total gravitational attraction now available, the final step
before starting the simulation is to set the vertical velocities of
the disk particles.  We integrate the vertical 1D Jeans equation
\citep[][eq.~422b]{BT08} for a slab, \ie\ neglecting radial
variations.  In this case
\begin{equation}
\sigma_{v_z}(R,z) = {1 \over \rho(R,z)} \int_z^\infty
  \rho(R,z^\prime) {\partial \Phi \over \partial z} \; dz^\prime,
\label{eq.vert1D}      
\end{equation}
where the vertical gradient of the total potential $\Phi$ is
determined numerically from the active particles, as well as the rigid
mass components.  This formula yields an acceptable equilibrium when,
as here, the radial excursions of disk particles are small; we do not
need the St\"ackel approximation recommended for hotter disks by
\citet{SB16}.

This set-up procedure yields a model in near perfect balance with the
initial virial ratio being $0.4953$.

\begin{figure}
\begin{center}
\includegraphics[width=\hsize,angle=0]{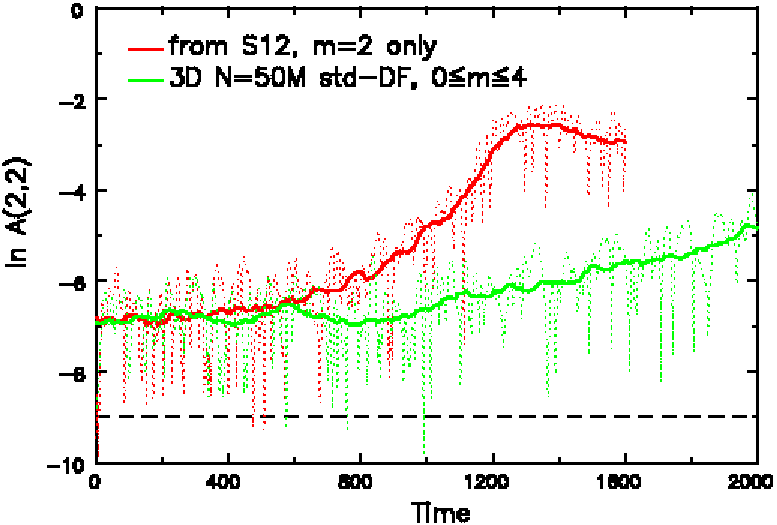}
\end{center}
\caption{Comparison of the amplitude evolution of the standard DF in
  2D (red) and 3D (green).  The red line is the $N=50$M case from S12,
  which used the standard DF with no groove.  The green line is from a
  model that uses the same DF but which was run with $0\leq m \leq 4$
  force terms on the 3D polar grid.  The amplitude rises in both
  simulations, though more slowly in 3D.}
\label{fig.std-DF}
\end{figure}

\subsection{Flaring}
As noted in the introduction, 3D motion allows for the possibility of
buckling instabilities that are caused by too large a disparity
between the radial and vertical velocity dispersions in the disk.  The
radial variations of the initial radial and vertical velocity
dispersions of our model are indicated by the black lines in the left
and middle panels of Fig.~\ref{fig.flare}, with the colored lines
being the same measurements at intervals of 6 time units from the
start revealing little change -- \ie\ the model is very close to
radial and vertical equilibrium.  The DF parameter $q=11.44$
(eq.~\ref{eq.DF}) sets $\sigma_R=0.284$ over the inner disk, which
decreases at larger $R$ because larger $E_{\rm rand}$ particles are
eliminated by the outer truncation -- see Fig.~\ref{fig.5654}.

The vertical velocity dispersion is set to support the adopted
constant initial disk thickness $z_0=0.3$ in the gravitational field
of the model (see eq.~\ref{eq.vert1D}), which causes the vertical
dispersion to decrease below 0.1 for $R\ga10$, and therefore to drop
below the threshold for buckling instability $3\sigma_{v_z} \la
\sigma_R$ \citet{Arak87}.  The consequence is that the outer disk
begins gradually to flare due to a local buckling instability.  The
modest increase in $\sigma_z$ for $R<20$, right panel, gradually
asymptotes to its final profile at $t \simeq 30$, and no further
increase was detectable for the remainder of the evolution.  It is
remarkable how small an increase in the vertical velocity dispersion,
middle panel of Fig.~\ref{fig.flare}, is required to relieve this
particular instability.

\subsection{Standard DF}
Fig.~\ref{fig.std-DF} compares the amplitude evolution of the $m=2$,
$\tan\gamma =2$ logarithmic spiral component in 2D and in 3D.  Both
runs start from the standard DF without a groove and manifest
secularly growing amplitude, which is slower in 3D, probably because
the gravity of disturbances is weakened by spreading the particles
vertically.  However, the qualitative evolution appears to be
unchanged by allowing 3D motion and including multiple secular
harmonics in the disturbance forces.

\subsection{Eliminating low-$E_{\rm rand}$ particles}
\begin{figure}
\begin{center}
\includegraphics[width=\hsize,angle=0]{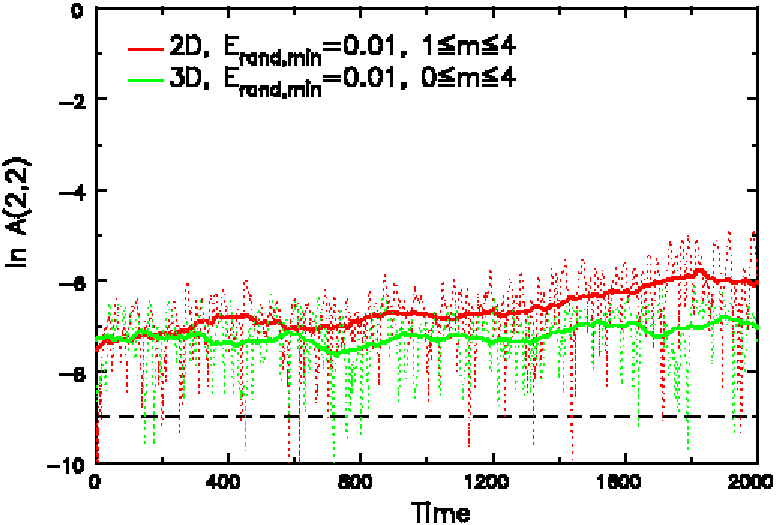}
\end{center}
\caption{As for Fig.~\ref{fig.std-DF} with $N=50$M, but for the
  modified DF with particles having $E_{\rm rand}< 0.01$ eliminated.
  The red line is from the same simulation as the green line in
  Fig.~\ref{fig.stable}, but continued to later times.}
\label{fig.Erand-min}
\end{figure}

Fig.~\ref{fig.stable} showed that the DF in Fig.~\ref{fig.DF5682}
stopped secular growth in 2D simulations.  Fig.~\ref{fig.Erand-min}
confirms that this result is unchanged when 3D motion is allowed.
Although we stopped the 3D simulation in Fig.~\ref{fig.std-DF} before
the non-axisymmetric disturbances had saturated, the steady rise of
the green curve there contrasts strongly with the flat behavior of the
green curve in Fig.~\ref{fig.Erand-min}, indicating a clear difference
between stable and mildly unstable cases.

\begin{figure}
\begin{center}
\includegraphics[width=.8\hsize,angle=0]{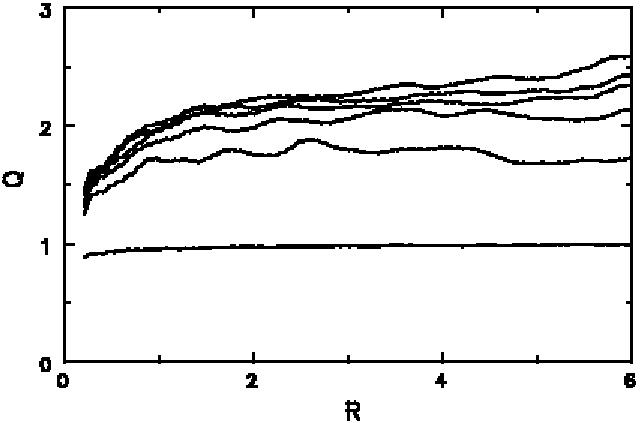}
\includegraphics[width=.8\hsize,angle=0]{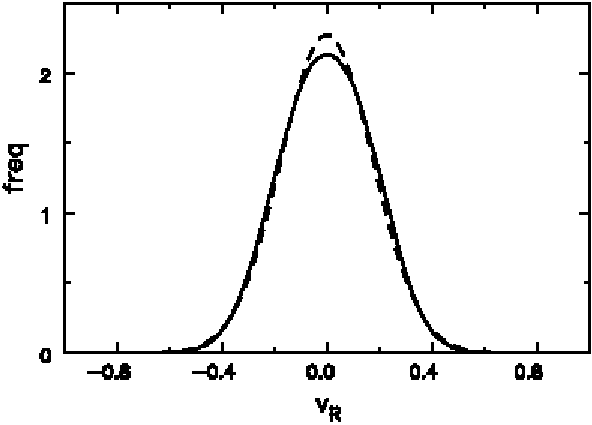}
\includegraphics[width=.8\hsize,angle=0]{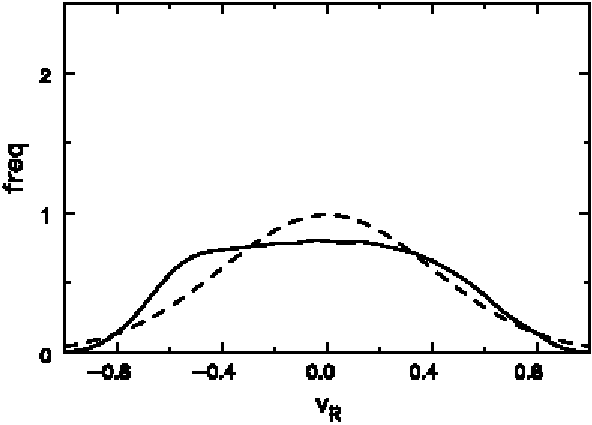}
\end{center}
\caption{Top: The evolution of $Q$ taken from a 2D, $N=2$M simulation
  of an SC disk model from \citet{SC14}.  The lines show the radial
  variation of $Q$ at times 0(200)1000 time units. \hfil\break Middle:
  The solid curve shows the distribution of radial velocities of all
  $\sim 950$K particles in the disk within the annulus $1 \leq R \leq
  4$ at $t=1000$. The dashed curve shows a Gaussian distribution
  having the same mean and $\sigma$. Notice that the velocity
  distribution is not quite as peaked as a Gaussian. \hfil\break
  Bottom: A similar plot to the middle panel, but from the region
  around the ILR of simulation that supported a strong spiral.  The
  spiral was triggered by a groove in the Mestel disk and was reported
  in \citep{SC22}.}
\label{fig.kurtos}
\end{figure}

\section{Discussion} \label{sec.discuss}
The sequence of events, described in \S\ref{sec.intro}, that gives
rise to secular growth in simulations with the standard DF is closely
related to the recurrent cycle of groove modes proposed by
\citet{SC19} to account for the transient spirals that develop
spontaneously in simulations of bar-stable disks.  A groove causes
exponential growth of a spiral which stores the angular momentum
extracted from the disk as its amplitude grows.  As the mode decays
the $L_z$ stored in the spiral drains onto the Lindblad resonances
\citep{SC22}, which carves new grooves at some radial distance from
the original.  The new grooves cause more spiral instabilities,
increasing random motion throughout the disk, causing activity to
subside in the absence of cooling by gas \citep{SC84}.

Thus gas-free disks are eventually stabilized when $Q$ has risen to a
value that weakens collective spiral instabilities, typically when $Q
\ga 2$, as shown by the time convergence of the lines in the top panel
of Fig.~\ref{fig.kurtos}.  The middle panel shows the distribution of
radial velocities in the disk at the last moment, which is less
strongly peaked than a Gaussian.  The velocity distribution presented
in the bottom panel resulted from scattering at the ILR of the strong
spiral described in the paper by \citet{SC22}.  This more dramatically
non-Gaussian distribution is that at $t = 360$, for which the
distribution in action space was illustrated in Fig.~4 of their paper.
These lower two panels reveal that spiral scattering depletes near
circular orbits, which is particularly pronounced for a strong spiral.
While the depletion is not total, as it was for the DF
(Fig.~\ref{fig.DF5682}) used in Figs.~\ref{fig.stable} and
\ref{fig.Erand-min}, we suspect it will have a reduced, but similar,
stabilizing effect.

\section{Conclusions} \label{sec.concls}
Previous work by \citet{Toom81}, described in \S\ref{sec.intro}, had
found that the half-mass Mestel disk having a gentle central cutout
and outer taper, possessed no small-amplitude instabilities, and
was therefore expected to be stable.  \citet{Sell12} confirmed, in
large-$N$ simulations that there were no {\em linear} instabilities,
but found unexpectedly that his simulations always manifested secular
growth by non-linear resonant scattering, described in \S1, which
undermined Toomre's prediction of disk stability.

Our present study (\S\ref{sec.groove}) of groove instabilities
confirmed our earlier suspicions that instability is provoked more
strongly by removing particles of low-$E_{\rm rand}$, than by those
having more eccentric orbits.  Since the secular rise of amplified
noise is driven by grooves/scratches in the DF, we suspected that
removing all particles having small $E_{\rm rand}$ from the DF might
inhibit the secular growth reported by \citet{Sell12}, and indeed it
did, as we found in Figs~\ref{fig.Ermin} and \ref{fig.stable}.

Here we extend our study of the instabilities of the Mestel disk to 3D
for the first time.  While achieving an equilibrium model from which
to start our simulations was hard, we found that motion in 3D made
little difference to the in-plane behavior.  The outer edge of the
tapered and truncated disk did support a mild buckling instability,
however, which ran its course quickly leaving the outer disk mildly
flared, but otherwise little changed.

Lack of growth in simulations is not, of course, proof of absolute
stability -- we can merely place an upper bound on the growth rate of
non-axisymmetric disturbances.  Since the orbit period at $r=6$ is
$\sim 37$ time units, the duration of the simulations in
Fig.\ref{fig.Erand-min} is over 50 orbits.  So it seems reasonable to
conclude that secular growth in our modified Mestel disk model is
negligible.

We note that our study has excluded a possible gas component within
the disk.  This neglect is appropriate, however, because we are
addressing the stability of featureless disks, such as those in S0
galaxies, which are believed to have very low gas content.

The rigid halo in our models is an important contributor to disk
stability and we have not here tested whether replacing it with mobile
particles would destabilize the entire model.  However, in a direct
test of this issue, \citet{Sell21} concluded that spiral disturbances
in a disk do not couple to a halo composed of mobile particles.
However, we caution the reader that if there were a mild bar
instability in a disk embedded in a rigid halo, a responsive halo may
well increase the growth rate of the instability.

One might worry that our finding that secular growth can be inhibited
by removing low-$E_{\rm rand}$ particles may have little relevance to
the stability of real galaxies, since it requires a DF that is
non-Gaussian in radial velocities (Fig.~\ref{fig.DF5682}).  However,
the DF of a galaxy disk is sculpted over time by resonant scattering,
which depletes the density of low $E_{\rm rand}$
particles. \citet{SC84} argued that star formation to replenish low
$E_{\rm rand}$ stars was essential to maintain spiral activity in a
disk, whereas here we are using a similar argument that a disk galaxy
can be stable if there are few low $E_{\rm rand}$ stars -- they having
been depleted by earlier spiral activity, that has now abated when the
supply of fresh gas has been removed. We show in Fig.~\ref{fig.kurtos}
that the spiral scattering does indeed introduce negative kurtosis
into the velocity distribution, which can be quite pronounced in the
case of strong spirals, and that change will inevitably add to the
stability of the entire disk.

\begin{acknowledgments}
We thank an anonymous referee for a careful read of the paper and
an insightful report.
JAS acknowledges the continuing hospitality and support of Steward
Observatory.
\end{acknowledgments}



\end{document}